\documentclass[sigconf]{acmart}
\usepackage{multirow,balance}

\AtBeginDocument{%
  }

\setcopyright{acmcopyright}
\copyrightyear{2023}
\acmYear{2023}
\acmDOI{XXXXXXX.XXXXXXX}

\acmConference[BIOKDD'23]{22nd International Workshop on Data Mining in Bioinformatics.}{August 9, 2023}{Long Beach, CA, USA}
\acmPrice{15.00}
\acmISBN{978-1-4503-XXXX-X/18/06}

\begin{document}

\title{DeepGATGO: A Hierarchical Pretraining-Based Graph-Attention Model for Automatic Protein Function Prediction}

\author{Zihao Li}
\affiliation{%
	\institution{College of Computer Science and Software Engineering}
	\institution{Shenzhen University}
	\city{Shenzhen}
	\country{China}}
\email{2110276126@email.szu.edu.cn}

\author{Changkun Jiang}
\authornote{Corresponding author.}
\affiliation{%
	 \institution{College of Computer Science and Software Engineering}
  \institution{Shenzhen University}
  \city{Shenzhen}
  \country{China}
}
\email{ckjiang@szu.edu.cn}

\author{Jianqiang Li}
\affiliation{%
\institution{National Engineering Lab for Big Data System Computing Technology}
  \institution{Shenzhen University}
  \city{Shenzhen}
  \country{China}
}
\email{lijq@szu.edu.cn}

\renewcommand{\shortauthors}{Li et al.}

\begin{abstract}
Automatic protein function prediction (AFP) is classified as a large-scale multi-label classification problem aimed at automating protein enrichment analysis to eliminate the current reliance on labor-intensive wet-lab methods. Currently, popular methods primarily combine protein-related information and Gene Ontology (GO) terms to generate final functional predictions. For example, protein sequences, structural information, and protein-protein interaction networks are integrated as prior knowledge to fuse with GO term embeddings and generate the ultimate prediction results. However, these methods are limited by the difficulty in obtaining structural information or network topology information, as well as the accuracy of such data. Therefore, more and more methods that only use protein sequences for protein function prediction have been proposed, which is a more reliable and computationally cheaper approach. However, the existing methods fail to fully extract feature information from protein sequences or label data because they do not adequately consider the intrinsic characteristics of the data itself. Therefore, we propose a sequence-based hierarchical prediction method, DeepGATGO, which processes protein sequences and GO term labels hierarchically, and utilizes graph attention networks (GATs) and contrastive learning for protein function prediction. Specifically, we compute embeddings of the sequence and label data using pre-trained models to reduce computational costs and improve the embedding accuracy. Then, we use GATs to dynamically extract the structural information of non-Euclidean data, and learn general features of the label dataset with contrastive learning by constructing positive and negative example samples. Experimental results demonstrate that our proposed model is no longer limited to specific biological populations and exhibits better scalability in GO term enrichment analysis on large-scale datasets.
\end{abstract}





\maketitle

\section{Introduction}
Proteins are large biomolecules or macromolecules that contain one or more long chains of amino acid residues and play a wide range of functions in living organisms. Each protein has one or more functions, including catalyzing metabolic reactions \cite{catalyzing}, DNA replication \cite{DNA_replication}, providing structure for cells and organisms, etc. In the past few decades, next-generation sequencing technologies \cite{next_generation} have enabled explosive the growth of protein sequence data due to their advantages in high throughput, scalability, and speed. However, the method of determining protein function through wet experiments wastes a lot of manual resources and time. The availability of various biological data such as subcellular localization, protein sequences, protein structures \cite{protein_structure}, and protein-protein interaction networks \cite{PPI} has increased the diversity of protein datasets, making it possible for automatic protein function prediction (AFP) via well-designed computational methods \cite{AFP}.

Existing computational methods like NetGO \cite{NetGO} and DeepFRI \cite{DeepFRI} integrate additional protein network data or structural data with the sequence information to enhance protein function prediction. However, due to the existence of data incompleteness and inaccuracy, as well as the reliance on the conservative nature of homologous sequences, these methods may exhibit suboptimal performance and high complexity \cite{limitation}. Since the differences in protein functions mainly lie in the arrangement of amino acid sequences, structural information and folding patterns are implicitly represented in the sequence features. Therefore, relying solely on sequence information for AFP has become a more realistic method. However, it is challenging to accurately characterize the sequence information due to the highly complicated interdependencies between the sequence data. 
Moreover, the current AFP approaches often rely on the curated Gene Ontology \cite{GO} (GO) data as functional annotation labels. Due to the inherent differences between sequence data and GO data, it is not feasible to simply combine them for feature extraction. Therefore, a popular approach in current research is to hierarchically process sequence and GO data to analyze the relationship between sequences and functional descriptions.


As the largest source of gene function system representation worldwide, GO is commonly used for protein enrichment analysis \cite{enrichment}.  Moreover, the widely used general protein knowledge base, UniProt \cite{Uniprot}, incorporates GO terms as part of the functional annotation scores for proteins. As a result, many computational methods focus on improving the accuracy and scalability of GO annotations as an evaluation metric for protein function prediction methods. GO annotation is created by associating genes or gene products with GO terms. These statements together constitute a ``snapshot'' of current biological knowledge. A GO annotation can be uniquely identified by four pieces of information: Gene product (maybe a protein, RNA, etc.), GO term, Reference, and Evidence. GO data contains three biological domains \cite{GO_2}: Molecular Function Ontology (MFO), Biological Process Ontology (BPO), and Cellular Component Ontology (CCO), and there are already over 44,000 concepts in 2021. The description of the GO structure can be viewed as a directed acyclic graph (DAG) with a root node, where each GO term represents a node and the relationship between terms can be viewed as edges in the graph. GO terms strictly follow the ``parent-child'' relationship, where child terms often have more specialized functions than parent terms, which is a reflection of metabolic processes in biosynthesis. As GO term datasets are not inherently flat labels, meaning that they do not possess the translational invariance like image or text data, when dealing with GO data, we must consider its irregular graph structure. Therefore, developing an AFP method that can capture the global GO graph structure information is an important problem that needs to be further addressed.

Recently, there have been methods based on CNNs to extract the relationships between GO terms (e.g., TALE \cite{TALE}). 
However, CNNs are not well-suited for handling non-Euclidean data. Moreover, this approach integrates the sequence information and GO graph embedding to extract features, and thus may not effectively capture the internal structural relationships between GO graph nodes. As a typical extension of GNNs \cite{GNN} methods for dealing with irregular graph structures, GCNs \cite{GCN} have been 
used to characterize GO data. In particular, PANDA2 \cite{PANDA2} 
takes GO terms and their edge connections as inputs and aggregates information from nodes and edges to generate new representations for GO nodes. However, this approach determines edge weights during the matrix construction phase, requiring prior knowledge of the graph's structure, thus significantly reducing flexibility. To address this issue, we resort to the Graph Attention Network \cite{GAT} (GAT), which introduces the attention mechanism into GNNs and provides a new solution for handling GO graph data. 
A GAT dynamically computes the ``degree of connectivity'' between nodes through point-wise operations, allowing for greater flexibility without the need for prior knowledge of the specific graph structure. However, considering only the GO graph structure is not sufficient because GO terms, as descriptions of protein functions, also contain \emph{semantic information} about the terms themselves. Therefore,  how to simultaneously capture the structure and semantic information of existing rigorously curated GO terms, while also ensuring broad coverage of our protein data annotations as much as possible is still a challenge to be addressed.

Since the fact that the GO graph is a data structure formed by connecting GO terms through graph structure, we abandon the idea of simple joint processing and adopt a \emph{hierarchical} approach that separately considers the structural and semantic information. Currently, most existing methods only compute one-hot embeddings of GO terms as node features or extract node features using multi-layer CNNs. However, these methods only focus on the features of individual nodes or local feature extraction. With the emergence of large-scale pre-trained models, we can efficiently extract feature embeddings of the labels using pre-trained model parameters. However, extracting feature embeddings solely from the labels is not sufficient, as there are still a significant number of proteins with incomplete GO term annotations. To address this issue, we resort to contrastive learning \cite{contrastive_learning} based on the idea of data augmentation, which can generate another version of essentially the same data and learn the general features of the dataset by comparing the two versions. This provides a new solution to the problem of incomplete coverage of labels and can be properly leveraged in our design.

To address the aforementioned challenges, we propose a sequence-based AFP method called DeepGATGO, which can improve the universality of low sequence homology and rare functional annotations. Considering the structural differences between sequence data and GO data, we cannot combine the two types of data from the start, so we adopt a hierarchical approach to handle these two types of data separately. More specifically, we first employ the ESM-1b \cite{ESM-1b} and BioBert \cite{BioBert} pre-trained language models to compute embeddings for protein sequences and GO terms, respectively. The adjacency matrix is then constructed based on the connection relationships among GO terms, which is used as input to the GAT layer. The GAT layer dynamically aggregates high-order representations within neighboring nodes using an attention mechanism, effectively addressing the limitations of previous methods that treated GO terms as flat labels or predetermined edge weights (thus neglecting the inherent structure of the GO graph). Subsequently, we utilize contrastive learning to construct positive and negative examples of GO terms and learn their general semantic features, addressing the issue of incomplete coverage of GO terms for all protein data. The main contributions of this work can be summarized as follows.
\begin{itemize}

	\item \emph{New Perspectives for the AFP Problem}: We outline the issues of the existing AFP methods in handling the sequence information and GO data. To address these issues, we propose a hierarchical learning approach that aims to characterize the relationship between sequences and functional descriptions and to capture both structural and semantic information of the GO graph. Such a solution approach can better exploit the sequence information and GO data for the AFP task.
	\item \emph{DeepGATGO Design}: We develop DeepGATGO, a sequence-based hierarchical AFP method that uses pre-trained models and GATs to focus on the fusion of sequence and label information. Moreover, we leverage contrastive learning to solve the problem of incomplete coverage of labels in our design. 
	\item \emph{Extensive Experiments}: We demonstrate the performance of our model on the Critical Assessment of Function Annotation 3 (CAFA3) dataset for protein function annotation and the more complex TALE dataset. The experimental results  validate the effectiveness of incorporating attention mechanisms in the extraction of GO label features, and verify the superior performance of our model for the AFP task. 
\end{itemize}

The rest of the paper is organized as follows. In Section \ref{related_work}, we summarize the related work. In Section \ref{PM}, we introduce the proposed method DeepGATGO, including the CAFA3 dataset and the detailed network modules. In Section \ref{er}, we implement the proposed method and compare our method with the state-of-the-art baseline methods. Finally, we conclude and discuss future work in Section \ref{con}.

\begin{figure*}[!t]
\centering
\includegraphics[width=0.95\linewidth]{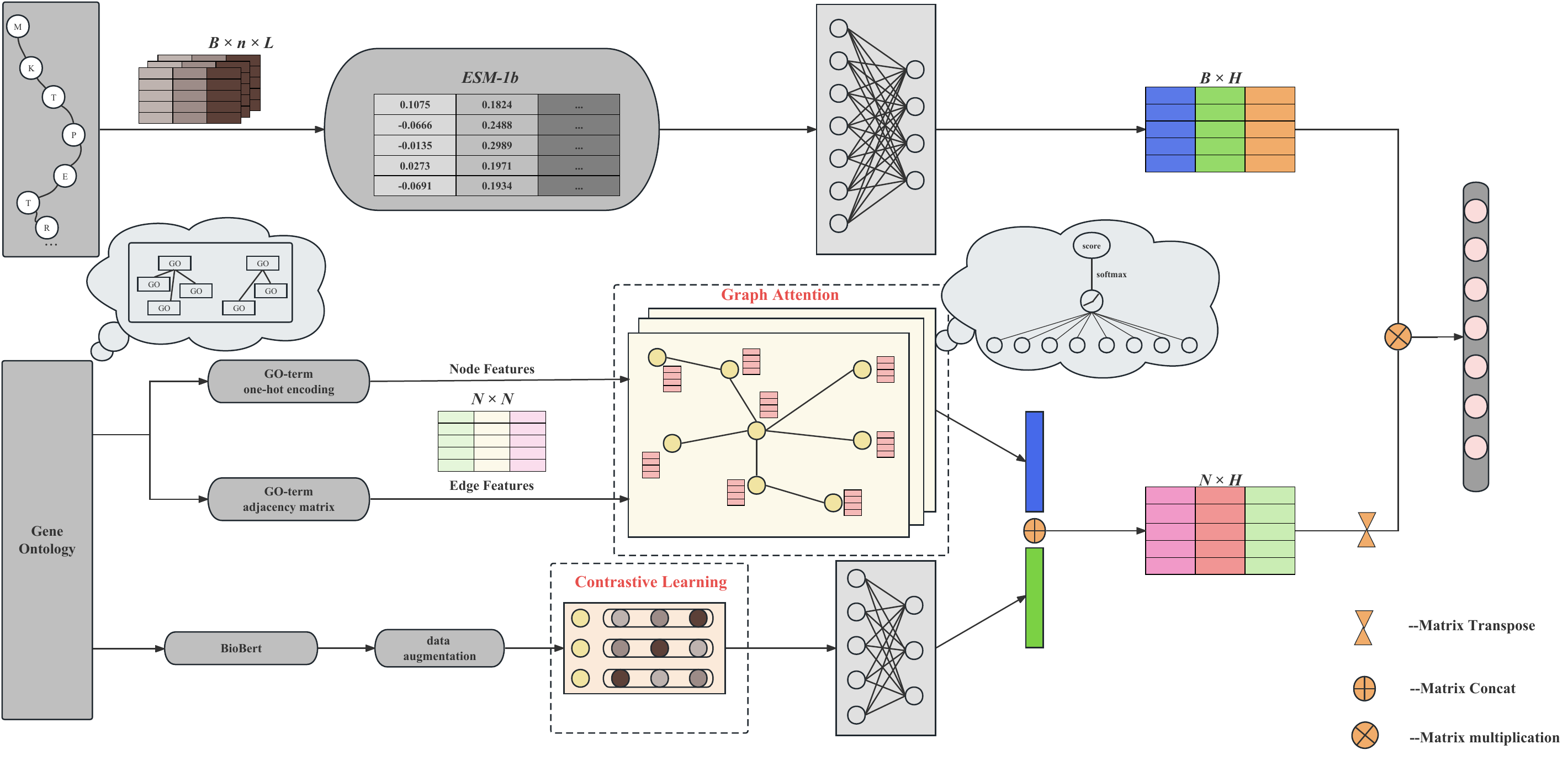}
\caption{A schematic procedure of DeepGATGO. The input includes protein sequences downloaded from UniProt and GO term labels downloaded from the CAFA3 benchmark dataset. Protein sequences are first embedded into sequence embeddings using the ESM-1b pre-trained model, while GO terms are embedded into semantic representations using the BioBert pre-trained model. The GO term feature vectors are then fed into two layers of graph attention (GAT) to capture structural information. The semantic information is captured using a contrastive learning layer. The structural and semantic features are combined with the sequence embeddings through fully connected layers, and the final protein function prediction is computed.}
\label{fig:label_example}
\end{figure*}

\section{Related Work} \label{related_work}
With the development of deep learning, there are more and more methods that use network-based approaches for AFP. Next, we will go over a series of typical studies among them. 

Initially, GO terms are used with random walk algorithms \cite{random-walk,NETSIM2} to search for probabilistic relationships in the GO graph. In particular, NETSIM2 \cite{NETSIM2} is based on a restart random walk algorithm that considers the global structure of the co-functional network. This involves randomly initializing and building random walkers across multiple nodes. However, due to the strict term inheritance relationships in the GO graph and the randomness of the initialization, this algorithm fails to consider each GO term's internal connections with its neighboring nodes fairly. 

Subsequently, DeepGO \cite{DeepGO} aims to predict GO labels based on traditional CNN methods from protein sequence and protein-protein interaction network \cite{PPI} (PPI) data. However, this method fails to consider the irregular graph structure of GO data itself. Simply embedding GO terms into a vector space does not effectively capture the structure and semantic information of the GO graph. Subsequently, Bourgeais et al. propose Deep-GONet \cite{Deep-GONet}, a self-explanatory deep learning model that integrates the GO label data into the neural network's hierarchical structure through a fully connected layer. The method takes GO terms as inputs, and embeds the biological knowledge contained in GO into a vector space through seven fully connected layers.

Furthermore, it is true that integrating protein network information can enrich the diversity within the model for AFP, but at the same time also increase the difficulty of training the model. In particular, You et al. propose a web server called NetGO \cite{NetGO} to improve the AFP performance by integrating a large amount of protein network information. NetGO ranks existing computational methods, effectively integrates protein sequence and network information, and utilizes a large amount of network information from the STRING database \cite{STRING} of all species to annotate proteins that have not been previously annotated through homologous transfer.

Due to the introduction of Transformer \cite{Transformer}, the current state of using traditional neural networks for predicting protein function has changed. The Transformer model introduces the attention mechanism and encoder-decoder module with a feed-forward neural network, which reduces the distance between any two tokens in the sequence to a constant and has good parallelism. In particular, TALE \cite{TALE} is based on the Transformer model and uses joint sequence-label embedding for protein function annotation. This method first encodes the sequence and feeds it to the Transformer-encoder module to extract hidden relationships between sequences, and then integrates label information into the output of the encoder module. Then, it further extracts internal features through convolutional and pooling layers to achieve protein function prediction. However, this method does not fully consider the internal relationships between GO term labels, but it lays a solid foundation for later AFP works. GCL-GO \cite{GCL-GO} is a method for protein function prediction using large-scale pre-trained models and graph neural networks. This method separately inputs protein sequences and GO term labels into pre-trained models ESM-1b \cite{ESM-1b} and BioBert \cite{BioBert}, and then uses graph convolutional networks (GCN) to extract hidden relationships between GO labels. However, GCN is a type of graph neural network that uses fixed node and edge relationships as inputs, and once the edge weight is input, it does not change.

Inspired by the above work, our proposed method leverages pre-trained language models to extract embeddings for both protein sequences and GO terms. Additionally, we utilize graph attention networks and contrastive learning to simultaneously capture the structural and semantic information of the GO graph.

\section{Proposed Method} \label{PM}
As mentioned above, we propose a hierarchical pretraining-based graph-attention  model for the AFP task. Figure 1 shows the overall flowchart of our model, which includes two types of inputs: protein sequences and GO term labels. It is worth noting that we use only these two types of data to construct the network architecture and generate the final prediction.

\begin{table}[]
\caption{Number of Sequences and Protein Annotations in our Datasets}
\centering
\begin{tabular}{|c||cccc|}
\hline
Dataset                & Statistics           & MFO    & BPO    & CCO    \\
\hline
\multirow{4}{*}{CAFA3} & Seq in Training Set   & 28,068 & 40,649 & 39,461 \\
                       & Seq in Validation Set & 7,017  & 10,163 & 9,866  \\
                       & Seq in Test Set       & 1,101  & 2,145  & 1,097  \\
                       & Number of GO terms    & 10,236 & 28,678 & 3,905  \\
\hline                       
\multirow{3}{*}{TALE}  & Seq in Training Set   & 38,323 & 54,206 & 52,257 \\
                       & Seq in Test Set       & 1,916  & 2,836  & 2,084  \\
                       & Number of GO terms    & 6,381  & 19,939 & 2,574  \\
\hline
\end{tabular}
\end{table}

\subsection{Datasets}
We use the CAFA3 dataset \cite{CAFA3}, which is introduced in the international protein function prediction competition, to conduct a large-scale evaluation of computational methods specifically designed for protein function prediction. Table 1 shows the detailed datasets we adopted. 
In particular, 
sub-terms in the datasets represent a further refinement of the parent term's function. Specifically, if a protein is annotated with a certain GO term, it will also be annotated with the parent terms of that GO term, in accordance with the hierarchical relationships of GO terms mentioned earlier. We establish the edge relationships of GO terms based on the "is a" and "part of" relationships in the CAFA dataset. Moreover, we include the dataset of the TALE method \cite{TALE} (which also utilizes the CAFA3 dataset) for comparison in our experimental section. All these combinations form two annotation datasets, as shown in Table 1.

\subsection{Pre-trained Model}

We leverage the pre-trained models ESM-1b and BioBert to efficiently extract feature embeddings of the labels. Specifically, ESM-1b takes protein sequences as input and aims to generate a large-scale, universally applicable pre-trained model for proteins through a high-capacity Transformer. After training, ESM-1b outputs information such as protein structure and homology in the form of features that are implicitly encoded within a single protein sequence. During training, ESM-1b uses a 33-layer model with 650 million parameters, and the input protein sequence is embedded into a 1280-dimensional residual-level embedding. The embedded matrix of the protein sequence that has undergone ESM-1b is then fed into a fully connected layer, which outputs sequence embedding features relevant to protein function.

BioBert is a biomedical language representation model designed to process difficult biomedical vocabulary using the Bert model. Thanks to the descriptions provided by Gene Ontology (GO) terms, we input the name and definition as a single statement into BioBert, and the semantic information output is used as input to the graph contrastive learning layer.

\subsection{Multi-head Graph Attention Network}
We employ a graph encoder based on Graph Attention Network (GAT) to obtain the hidden relationships between a single GO term and its neighboring GO terms. GAT is a graph neural network that utilizes attention mechanisms, which allows it to fully capture the interactions between specified nodes and their adjacent nodes, a feature that is lacking in traditional neural networks and Graph Convolutional Networks (GCN). Figure 2 shows the implementation of our attention mechanism.

\begin{figure}
	\centerline{\includegraphics[width= 80mm]{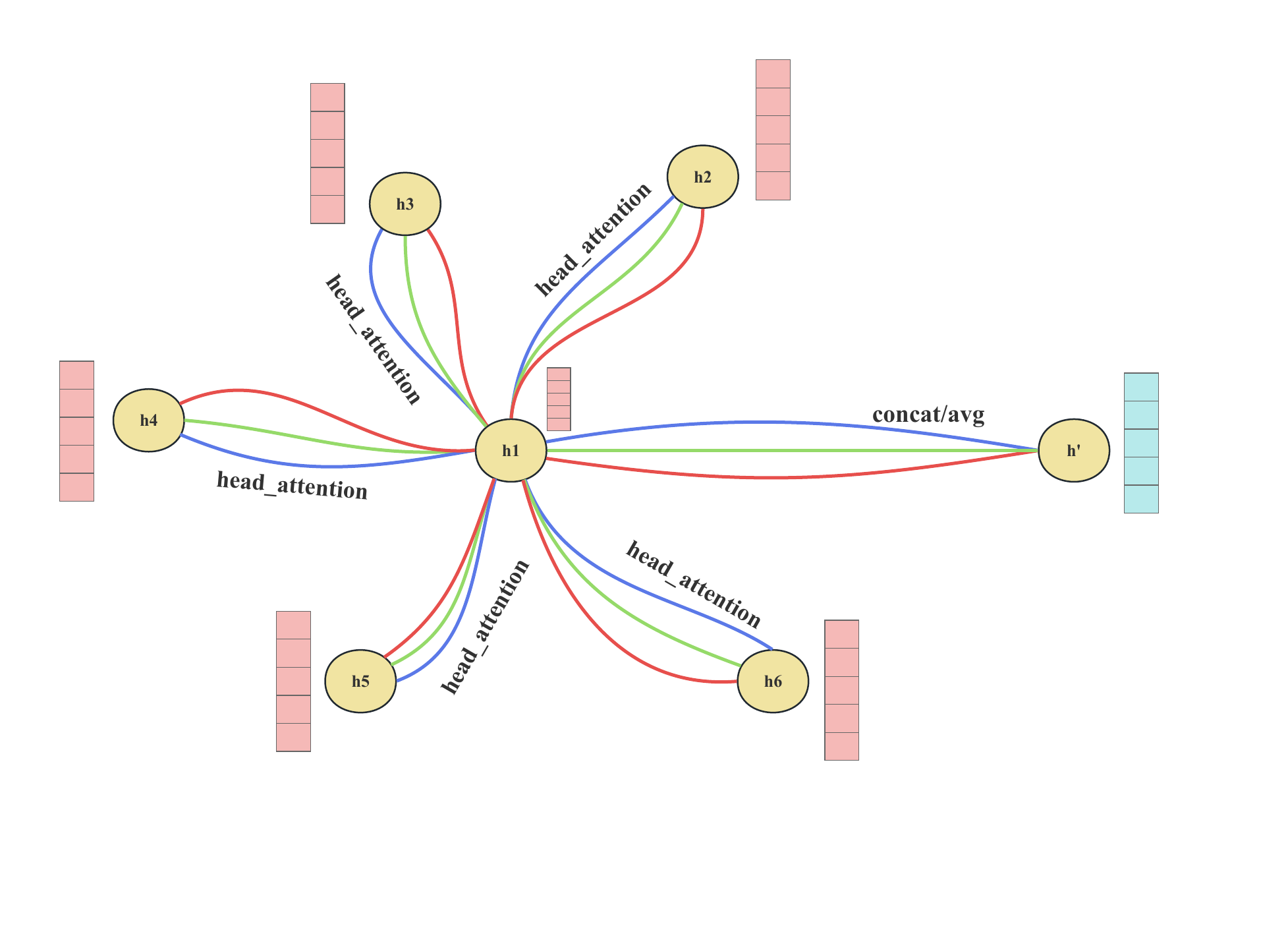}}
	\caption{The workflow of the graph attention layer involves treating GO terms as nodes in a graph structure and constructing an adjacency matrix for all GO terms connected to that node. Our network module uses multi-head attention, with different colors representing different heads' attention to the central node. With different heads, the central node $h_1$ can compute different embeddings, and we concatenate all the embeddings to generate the attention score $h'$ for that node.}
	\label{fig_2}
\end{figure}

We treat the GO terms as nodes in the GAT layer and use a one-hot encoding process to generate the initial node features $H_0$. We construct the adjacency matrix of the GO graph by setting edge weights to 1 for GO terms that have edges connecting them and 0 for those that do not. To reduce computational complexity and minimize training time, we embed the node features into a sparse matrix.

To fully capture the expressive power of GO terms and map input features to higher-level features, we initialize weight matrix $W$ and use it to map the input GO terms embeddings to the desired output dimension. We apply parameterized shared linear transformations of $W$ to each node. Then, we use a multi-head attention mechanism on the nodes, where the node representations are transformed and then used to compute the attention coefficients. The attention coefficients are used to weight the importance of neighboring nodes in computing the output representation of each node:

\begin{equation}
    e_{ij} = \Phi(\textbf{W}\Vec{h_i}, \textbf{W}\Vec{h_j}),
\end{equation}
where $e_{ij}$ represents the "importance" of node $j$ to node $i$, and $\Phi$ is a mapping function used to transform two features into a single real number. In our model, all nodes connected by an edge are allowed to participate in the attention score computation with their directly connected nodes, while the edges between unconnected nodes are set to negative infinity. In this part, we focus only on the structural information between nodes in the graph, and we inject a masking mechanism into the model. We compute $e_{ij}$ for node $j \in N_i$, where $N_i$ is the first-order neighborhood of node $i$. To facilitate the comparison of coefficient results between nodes, we normalize the output results by inputting them into a softmax layer as follows.

\begin{equation}
    \alpha_{ij} = softmax_j(e_{ij}) = \frac{exp(e_{ij})}{\sum_{k\in N_i}exp(e_{ik})}.
\end{equation}
In our multi-head attention mechanism, each head uses the aforementioned computing method to compute the attention score, and the representations obtained by all heads are concatenated as the final score in the output. In our experiment, the attention mechanism is a feed-forward neural network that takes the output and passes it through a non-linear LeakyReLU layer. The final score can be represented as:

\begin{equation}
    \alpha_{ij} = \frac{exp(LeakReLu(\Vec{a}^T[\textbf{W}\Vec{h_i}||\textbf{W}\Vec{h_j}]))}{\sum_{k\in N_i}exp(LeakyReLu(\Vec{a}^T[\textbf{W}\Vec{h_i}||\textbf{W}\Vec{h_k}]))},
\end{equation}
where $\cdot^T$ denotes transpose, and $||$ denotes concatenation operation.

\subsection{Contrastive Learning}
After the graph attention layer, the GO terms have successfully extracted internal structural information. Then, we further extract semantic information features between GO terms through contrastive learning. Figure 3 shows the workflow of the contrastive learning module. Inspired by the idea of data augmentation, we use the shuffled feature vectors of the sentences processed by BioBert as negative samples, and the average of the first five connected GO term features as positive samples. The objective is to increase the size of the existing dataset and narrow the distance between the original samples and the positive samples during training and move away from the negative samples.

We input the extracted semantic information features of GO terms into a fully connected layer, and concatenate them with the structural scores computed in Section 3.3. Then, we perform a simple matrix multiplication between the concatenated result and the protein sequence features to compute the contribution of each protein sequence to each GO term, and apply a sigmoid operation to the result, defined as follows:

\begin{equation}
    Y = sigmoid(SH^T),
\end{equation}
where $H$ represents the concatenated GO feature matrix, and $S$ represents the protein sequence feature embedding matrix.

\begin{figure}
	\centerline{\includegraphics[width=\linewidth]{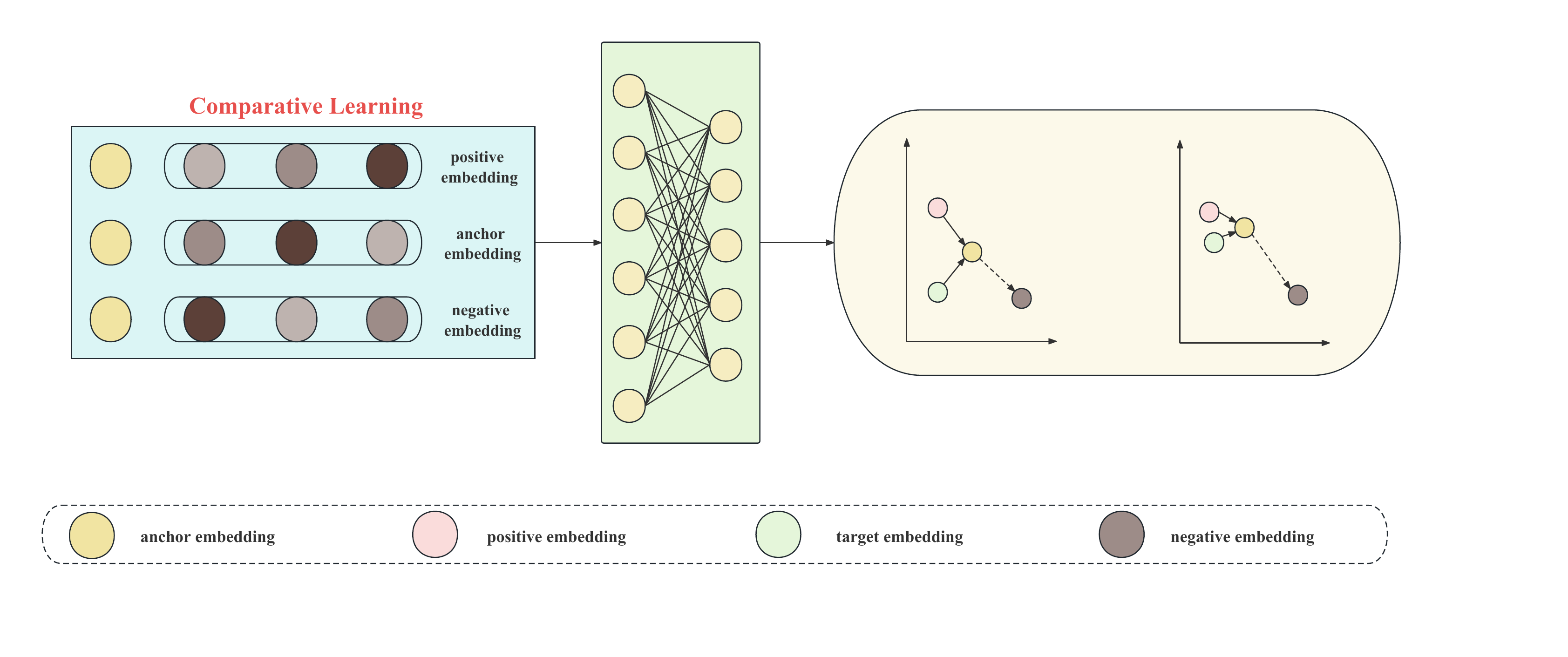}}
	\caption{The workflow of contrastive learning. We consider the GO term vectors computed by the BioBert pre-trained model as the raw data. We extract the first $n$ features of the vector, take their average as the positive sample for contrastive learning, and randomly shuffle the features of each vector to generate the negative sample.}
	\label{fig_3}
\end{figure}

\subsection{Loss Function}
In the choice of the loss function, as we compute the association between GO terms based on both structure and semantic information, and construct new positive and negative samples through graph contrastive learning, we finally choose triplet loss \cite{triplet_loss} as one of the loss functions in the training. Triplet loss takes a triplet as input, including an anchor sample, a positive sample, and a negative sample. When computing the structural loss, we use the structural information extracted by GAT as the positive sample. Similarly, when computing the semantic information loss, we use the samples with integrated neighbor information from contrastive learning as the positive sample. The specific formulas for the loss functions are given as follows.

\begin{equation}
    L_{St} = \frac{1}{N}\sum_{i\in N}\left\{d(h,h^+)^2 - d(h,h_i^-)^2 + \beta\right\},
\end{equation}
\begin{equation}
    L_{Se} = \frac{1}{N}\sum_{j\in N}\left\{d(h,\Tilde{h}^+)^2 - d(h,h_j^-)^2 + \beta\right\},
\end{equation}
where $L_{St}$ represents the loss of structural information, $L_{Se}$ represents the loss of semantic information, $h^+$ represents the structural information extracted by GAT, $\Tilde{h}^+$ represents the sample with integrated neighbor information, and $h^-$ represents the feature information of GO terms after random shuffling. The parameter $\beta$ represents the margin parameter in the triplet loss, which is used to prevent the positive and negative examples from being too close after training.

The second loss function we used during the training process is binary cross-entropy. Our final loss function can be expressed by

\begin{equation}\label{eq:loss}
    L = \gamma(\delta L_{St} + (1 - \delta)L_{Se}) + (1 - \gamma)L_{BCE},
\end{equation}
where $\gamma$ and $\delta$ are parameters we defined to control the preference of the loss function.

\section{Experimental Results} \label{er}
\subsection{Protein Sequence Dataset}
For the selection of protein sequences, we use the GCL-GO computing method as our baseline. The dataset for this method is downloaded from UniProtKB, which is currently the most widely used protein sequence database and has undergone careful curation. We filter the sequences based on the following criteria: (1) removal of sequences with lengths exceeding 1024 or containing non-standard amino acids, and (2) retention of sequences with high-quality functional annotations, i.e., filtering out sequences without biological data annotations. The resulting dataset is divided into training, validation, and testing sets.

\subsection{Baselines}
In order to better demonstrate the performance of our model, we have selected several widely accepted competitive methods for comparison. Among them, DeepGO \cite{DeepGO}, DeepGOPlus \cite{DeepGOPlus}, TALE \cite{TALE}, DeepGraphGO \cite{DeepGraphGO}, and GCL-GO \cite{GCL-GO} are all deep learning-based methods. It is worth mentioning that DeepGO and DeepGraphGO incorporate additional protein network information beyond just the protein sequences. DeepGO is a deep learning method that predicts protein functions by using multi-layer neural networks to analyze protein sequences and known protein interactions. TALE is a hierarchical perception method that combines sequence and label data information, and applies a transformer encoder to extract semantic information from sequences. TALE uses CNN and fully connected layers to extract feature information from joint embedding sequences and GO terms for prediction. DeepGraphGO and GCL-GO are two AFP methods that utilize graph neural networks. These two methods fully utilize the protein sequence and higher-order protein network information through GCN and different sequence processing methods. 

\subsection{Evaluation Metrics}
For the performance evaluation of our method DeepGATGO on the test set, we choose to use the following two evaluation metrics, i.e., $F_{max}$ and AUPRC. $F_{max}$ is the official evaluation metric proposed by CAFA, which combines precision and recall to measure the performance of protein function prediction. It is represented as below:

\begin{equation}
	F_{max} = \mathop{max}\limits_{t} \left( \frac{2 \times \overline{Pre}(t) \times \overline{Rec}(t)}{\overline{Pre}(t) + \overline{Rec}(t)} \right),
\end{equation}
where $\overline{Pre}(t)$ represents the proportion of true positive samples among the samples predicted as positive under the threshold $t$, while $\overline{Rec}(t)$ represents the proportion of samples predicted as positive among all samples that are actually positive under the threshold $t$. In order to facilitate our calculation, we iterate the threshold $t$ in the range of 0 to 1 with a step size of 0.01.


AUPRC is commonly used as an evaluation metric for model performance in binary classification tasks. It aims to measure the priority of positive samples ranked ahead of negative samples. In protein function prediction tasks, we use the AUPRC metric to evaluate the performance of predicting whether a protein has a specific function, which is particularly suitable for highly imbalanced data in such tasks. Therefore, we combine the final predicted label vector and use AUPRC as the evaluation criterion.

\begin{table}[]
	\caption{Performance comparison of our model and other competing methods on CAFA3 dataset}
	\begin{tabular}{|l|lll|lll|}
		\hline
		\multicolumn{1}{|c|}{\multirow{2}{*}{Method}} & \multicolumn{3}{c|}{Fmax} & \multicolumn{3}{c|}{AUPRC}                 
		\\ \cline{2-7} 
		\multicolumn{1}{|c|}{}& \multicolumn{1}{c|}{MFO} & \multicolumn{1}{c|}{BPO} & \multicolumn{1}{c|}{CCO} & \multicolumn{1}{c|}{MFO} & \multicolumn{1}{c|}{BPO} & \multicolumn{1}{c|}{CCO} \\ \hline
		DeepGO     & 0.534 & 0.384 & 0.550 & 0.412 & 0.245 & 0.451    \\
		DeepGOPlus & 0.587 & 0.504 & 0.656 & 0.514 & 0.376 & 0.604 \\
		DeepGraphGO & 0.607 & 0.518 & 0.667 & 0.558 & 0.352 & 0.657  \\
		TALE  & 0.467 & 0.418 & 0.637 & 0.360  & 0.283 & 0.610 \\
		GCL-GO & 0.610 & 0.515 & 0.674 & 0.562 & 0.395 & 0.669 \\
		\hline
		DeepGATGO & \textbf{0.617}& \textbf{0.528}& \textbf{0.679} & \textbf{0.568}& \textbf{0.406}& \textbf{0.673}  \\
		\hline
	\end{tabular}
\end{table}

\begin{table}[]
	\caption{Performance comparison of our model and other competing methods on TALE dataset}
	\begin{tabular}{|l|lll|lll|}
		\hline
		\multicolumn{1}{|c|}{\multirow{2}{*}{Method}} & \multicolumn{3}{c|}{Fmax} & \multicolumn{3}{c|}{AUPRC}                 
		\\ \cline{2-7} 
		\multicolumn{1}{|c|}{}& \multicolumn{1}{c|}{MFO} & \multicolumn{1}{c|}{BPO} & \multicolumn{1}{c|}{CCO} & \multicolumn{1}{c|}{MFO} & \multicolumn{1}{c|}{BPO} & \multicolumn{1}{c|}{CCO} \\ \hline
		DeepGO  &  0.432 &  0.260 &  0.587 &  0.355 &  0.201  &  0.516   \\
		DeepGOPlus & 0.634 & 0.384 & 0.632 & 0.587 & 0.235 & 0.578 \\
		DeepGraphGO & 0.601& 0.352 & 0.677 & 0.543 & 0.228 & 0.658\\
		TALE  & 0.578 & 0.336 & 0.658 & 0.514  & 0.247 & 0.635 \\
		GCL-GO & \textbf{0.636} & 0.385 & 0.683 & 0.612 & \textbf{0.287} & 0.671 \\
		\hline
		DeepGATGO & 0.627& \textbf{0.398}& \textbf{0.694} & \textbf{0.620}& 0.280& \textbf{0.683}  \\
		\hline
	\end{tabular}
\end{table}

\begin{figure*}[!t]
	\centering
	\includegraphics[width=\linewidth]{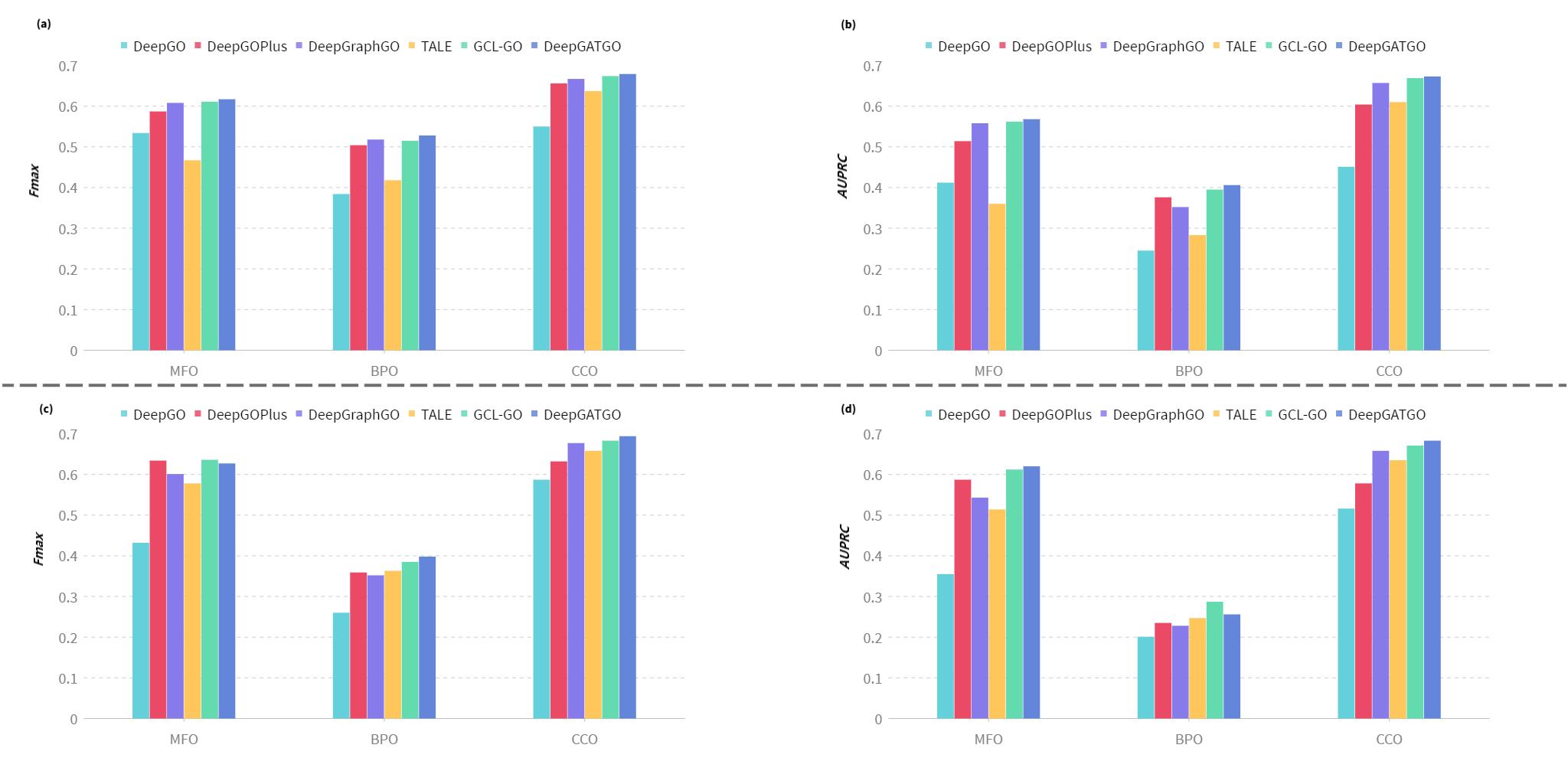}
	\caption{The performance of our model and competing methods on the CAFA3 and TALE datasets. (a) and (c) show the Fmax values of our model and the comparison methods on the two datasets, while (b) and (d) show the AUPRC value distributions on the two datasets.}
	\label{fig_5}
\end{figure*}

\subsection{Performance on the CAFA3 Dataset}
We train DeepGATGO separately on the MFO, BPO, and CCO datasets of GCL-GO. We use two multi-head graph attention layers because, in preliminary experiments, we find that increasing the number of GAT layers does not significantly improve our results, except for increasing the computational cost. We use the Adam optimizer and, like GAT, we use LeakyReLU as the activation function. To prevent overfitting, we apply dropout to the GAT layers and after the GAT layers. We train our models for 9 days on two NVIDIA GeForce RTX 3090 GPUs and report the average of the top three results as our evaluation metric.

Table 2 presents the performance of our proposed method, DeepGATGO, in predicting protein functions. We achieved satisfactory performance on the GCL-GO datasets, with Fmax scores of 0.617, 0.528, and 0.679 for MFO, BPO, and CCO, respectively. It is worth noting that we also achieved satisfactory AUPRC scores, with AUPRC scores of 0.568, 0.406, and 0.673 for MFO, BPO, and CCO, respectively. The experimental results clearly demonstrate the effectiveness of our model as it outperforms competing methods in all three biological domains. This achievement can be attributed to the utilization of the GAT layer, which allows for dynamic attention to the global features of the GO graph. Furthermore, the incorporation of contrastive learning enables our model to learn more generalized features of GO terms, contributing to its superior performance. It is worth mentioning that DeepGO and DeepGraphGO take into account both sequence information and network information of protein species. So our model achieves state-of-the-art performance even without utilizing additional protein network information, while also having lower computational costs.

\begin{figure}
	\centerline{\includegraphics[width=\linewidth]{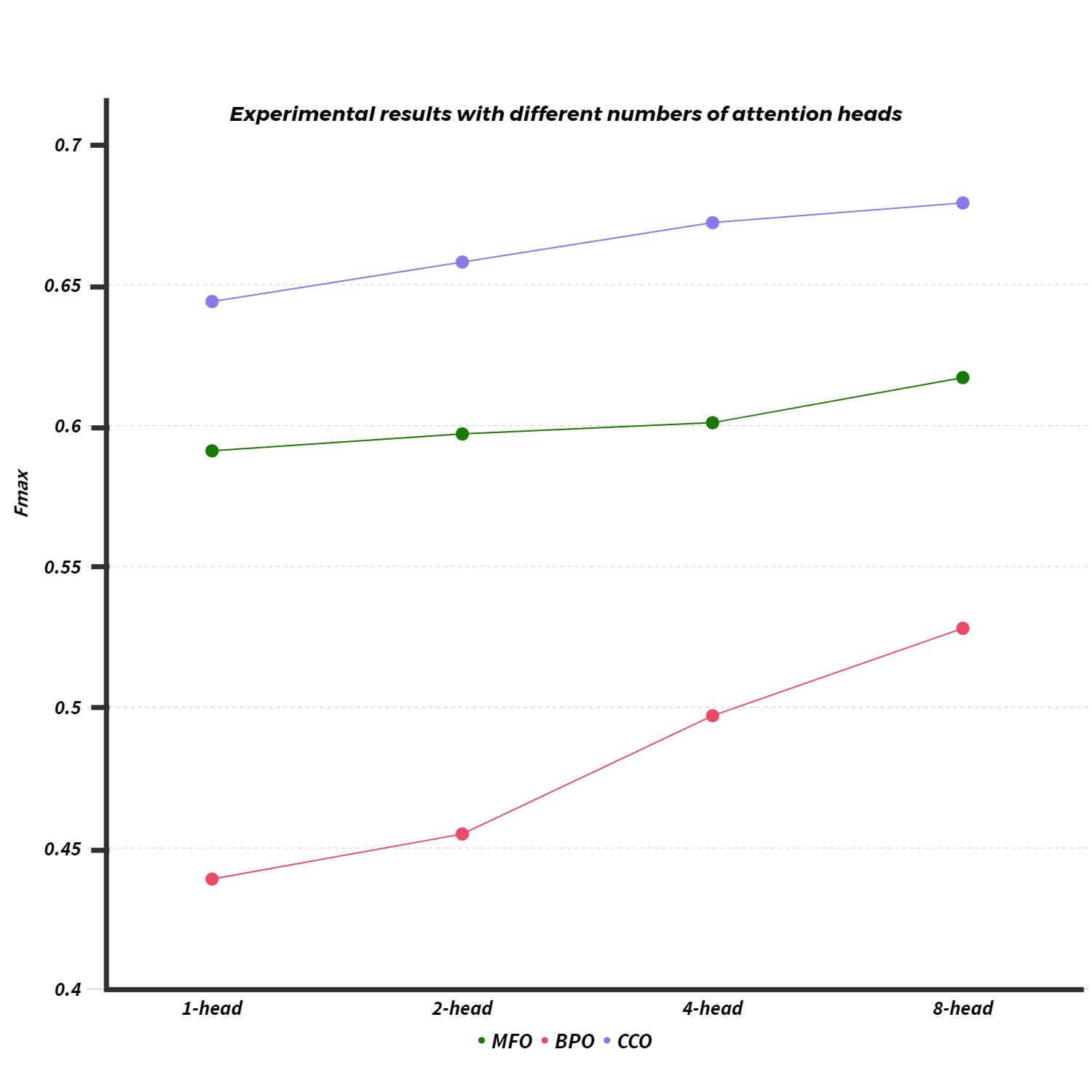}}
	\caption{The comparison of experimental results of our model with different numbers of "heads" selected in the graph attention (GAT) layer, and the performance of our model reached saturation after using 8 attention heads.}
	\label{fig_4}
\end{figure}

\subsection{Performance on the TALE Dataset}
We also compare the performance of DeepGATGO with the competing methods on the TALE test dataset. Table 3 shows the results. From Table 3, we can see that DeepGATGO demonstrates favorable performance on the dataset. However, it slightly underperforms competing methods in terms of the $F_{max}$ metric for Molecular Function (MFO) and the AUPRC metric for Biological Process (BPO). This discrepancy may be attributed to the limited number of GO terms included in TALE. Consequently, the advantages of dynamically capturing the global features of graph nodes may not be fully manifested in this context.

Moreover, Figure 4 shows the performances of all competing methods on the CAFA3 and TALE datasets, respectively. From Figure 4, we can see more clearly that while DeepGATGO does not outperform the state-of-the-art baseline methods in a few metrics, its performance on large-scale GO datasets is sufficient to demonstrate that our approach is more suitable for protein function annotation at a larger scale. This is attributed to the targeted extraction of internal features from GO data using graph attention networks and contrastive learning, as well as the reliability achieved by solely using protein sequences as the input information in our model.

\subsection{Performance on Different Attention Heads}
The experimental results mentioned above demonstrate the effectiveness of incorporating the GAT module in processing the GO graph. As an implementation of the attention mechanism in GNNs, GAT enhances the features of GO nodes using a shared parameter matrix $W$ and maps the concatenated high-dimensional features to attention scores. Based on the experimental experience in Transformer models, it has been shown that using multi-head attention in the attention module often yields better feature extraction performance than using a single head. This indicates that having more attention heads can more comprehensively capture the underlying relationships in the processed data.

To show the impact of attention heads on the prediction results, we evaluate the performance of our model with different numbers of attention "heads" in the graph attention layer. Figure 5 shows the prediction results of our model on different attention heads. As shown in Figure 5, our model's performance improves to some extent with an increasing number of "heads". However, considering the computational cost, we also attempt to use more attention heads, but the model's performance on MFO does not show significant improvement. In fact, there is even some degree of fluctuation in the experimental results for BPO and CCO. This may be attributed to the more complex structural relationships between GO terms in the MFO domain, which are difficult to capture. As a result, a smaller number of heads may not effectively capture the "attention" of each GO term towards other terms. On the other hand, using 8 attention heads has proven to be sufficient for our model to focus on the global structural features of the GO graph. Therefore, we can speculate that selecting an appropriate number of attention "heads" can optimize our model's performance to some extent.

\subsection{Performance under Different Values of $\delta$}

In the other aspect, considering the varying degrees of influence of GO data structure information and semantic information on the prediction results, we conduct comparative experiments to explore the impact of the parameter $\delta$ in Eq. (\ref{eq:loss}) on the performance of our final model. As shown in Table \ref{tab:delta}, our model tends to achieve optimal performance when $\delta$ is set to 0.5 or 0.6. This can be attributed to the more prominent impact of the GO graph structure information extracted by GAT layers during protein function prediction utilizing GAT layers and contrastive learning layers. Therefore, the selection of an appropriate method to further extract structural information from GO graph data becomes particularly crucial.

\begin{table}[]
\caption{Performance comparison of our model under different values of $\delta$ on CAFA3 dataset}\label{tab:delta}
\begin{tabular}{|c|ccc|ccc|}
\hline
\multirow{2}{*}{Different $\delta$} & \multicolumn{3}{c|}{Fmax}                                            & \multicolumn{3}{c|}{AUPRC}                                           \\ \cline{2-7} 
                         & \multicolumn{1}{c|}{MFO} & \multicolumn{1}{c|}{BPO} & CCO            & \multicolumn{1}{c|}{MFO} & \multicolumn{1}{c|}{BPO} & CCO            \\ \hline
$\delta$=0.7                    & 0.612                    & 0.526                    & 0.676          & 0.563                    & 0.399                    & 0.671          \\
$\delta$=0.6                    & \textbf{0.617}           & \textbf{0.528}           & \textbf{0.679} & \textbf{0.568}           & 0.404            & \textbf{0.673} \\
$\delta$=0.5                    & 0.616                    & \textbf{0.528}           & \textbf{0.679} & \textbf{0.568}           & \textbf{0.406}                    & 0.672          \\
$\delta$=0.4                    & 0.613                    & 0.526                    & 0.675          & 0.566                    & 0.403                    & 0.670          \\
$\delta$=0.3                    & 0.610                    & 0.524                    & 0.673          & 0.562                    & 0.401                    & 0.669          \\ \hline
\end{tabular}
\end{table}

\section{Conclusion} \label{con}
In this paper, we propose the design of DeepGATGO, a novel hierarchical pretraining-based graph-attention model for automatic protein function prediction. 
DeepGATGO only takes protein sequences as inputs and can achieve state-of-the-art performance in predicting three levels of Gene Ontology (GO). Our method obtains satisfactory results by only using sequence information due to the difficulty in obtaining accurate protein structure and binding site data. In particular, 
we use the pre-trained model ESM-1b to compute protein sequence embedding vectors. Moreover, we use the graph attention network to compute the structural information of GO terms, and use the pre-trained model BioBert to compute the semantic information of GO terms. To our knowledge, GO terms have strict parent-child relationships, making hierarchy-aware methods more suitable for our label data. The performance of DeepGATGO on both CAFA3 and TALE datasets outperforms previous sequence-based methods, meaning that DeepGATGO is no longer limited to specific biological species for predicting protein functions based on sequences. That is, DeepGATGO is a more generalized and better-performing model for automatic protein function prediction.

Given the current state where protein function annotation still relies heavily on labor-intensive wet-lab experiments, we aim to further propose a more generalized and scalable deep learning model based on our existing model. We hope that our model can not only predict protein functions accurately on highly annotated protein sequences but also on those with limited annotations. Moreover, we aim to expand our model to enrich the current available GO term annotations.

\begin{acks}
This work was supported by the National Key R\&D Program of China under Grant 2020YFA0908700, the National Natural Science Foundation of China under Grants 61902255 and 62006157, the Natural Science Foundation of Guangdong Province-Outstanding Youth Program under Grant 2019B151502018, as well as the Basic Research Project of Shenzhen Science and Technology Program under Grants JCYJ20190808163417094 and JCYJ20190806112215116.
\end{acks}

\balance
\bibliographystyle{unsrt}
\bibliography{conf}

%
%
%
%
%
%
%
%

\end{document}